# Faster AI, Uneven Frontier:

## Rapid Crossings, a Jagged Frontier, and the Repositioning of Human Judgment

Ancuta Margondai, Julie Rader, Emma Rader,

Sara Willox, and Mustapha Mouloua

University of Central Florida

**Abstract**

Between 2023 and 2026, frontier AI systems crossed documented human expert baselines on a growing set of bounded, well-specified, evaluable cognitive tasks, including graduate-level science questions, competition mathematics, software-engineering benchmarks, and structured diagnostic reasoning, while the length of tasks such systems can complete at 50% reliability doubled roughly every seven months. These crossings are rapid and broad, but the frontier is jagged: humans retain decisive advantages in long-horizon reliability, genuinely novel problems, calibrated self-knowledge, sample-efficient learning, and embodied action, and benchmark results overstate deployed capability for reasons that are themselves now documented, namely contamination, construct validity, vendor self-evaluation, and the gap between 50% reliability and the reliability that economic work requires. Concurrently, humans increasingly use these systems as cognitive extensions. The offloading literature predicts costs to unaided skill, and early field evidence is consistent with such costs, though the largest meta-analytic evidence on prior technologies points the other way, and the question of whether generative AI differs is open. Finally, the experimental record on human-AI collaboration shows that naive combination often underperforms the stronger partner, implying that the human contribution must be repositioned toward specification, verification, and oversight, a shift visible in experiments but, so far, barely visible in field labor-market data. This paper states the resulting position, rapid crossings on a jagged frontier with a human role that must be redesigned rather than defended, and draws out its theoretical and practical implications.

**Faster AI, Uneven Frontier: Rapid Crossings, a Jagged Frontier, and the Repositioning of Human Judgment**

## 1. Introduction

What was until recently an intuition is now a measured phenomenon. Generative AI has diffused faster than the personal computer or the internet at comparable stages: in the first nationally representative U.S. survey of its use, 39.4% of adults aged 18 to 64 reported using generative AI within two years of ChatGPT's release, nearly double the adoption rate of the personal computer three years after its introduction (Bick et al., 2025). By late 2025, 21% of U.S. workers reported that at least some of their work is done with AI, up from 16% a year earlier, and 31% of adults reported interacting with AI several times a day (Pew Research Center, 2026). Adaptation, meanwhile, lags in the same surveys: half of U.S. adults say the increased use of AI in daily life makes them more concerned than excited, up from 37% in 2021, while only 10% report the opposite (Pew Research Center, 2026). The sense that artificial intelligence is advancing faster than people and institutions can adapt, and that people now think *with* these systems rather than merely *use* them, is therefore no longer voiced only in workplaces, classrooms, and policy debates; it is visible in adoption and attitude data. This paper asks what the published record supports, including the published record that cuts against it.

The thesis presented here is intentionally more focused than the common one. It does not assert that "the human-AI gap is widening," a statement that is either incorrect on some metrics, difficult to prove as stated, or only true under conditions its supporters seldom specify. Instead, it argues that: (1) frontier AI has quickly surpassed documented human expert benchmarks on specific tasks that are bounded, well-defined, and measurable, with the pace of this advancement also quantified (Section 2), while the overall capability frontier remains uneven, with clear and lasting human advantages (Section 3); (2) AI increasingly acts as a cognitive extension, following the framework of Clark and Chalmers (1998), which makes questions about the costs of offloading vital, though this evidence is still debated (Section 4); and (3) experimental data suggests human roles should shift from focusing on throughput to emphasis on judgment, verification, and oversight, a shift demonstrated by experiments but not yet widely supported by field data (Sections 5 and 6). Section 7 explores the theoretical and practical consequences of maintaining all three claims and their nuances, while Section 8 highlights the boundaries of available evidence.

## 2. Rapid Crossings: What the Record Shows

### 2.1 The Rate of Progress, Stated Carefully

The most cited quantitative measure of progress comes from METR: the length of tasks (in human working time) that AI systems can complete at 50% reliability has doubled approximately every seven months since 2019, with possible acceleration in 2024 (Kwa et al., 2025), and broadly similar doubling rates appear across nine benchmark domains (METR, 2025). Two qualifications are required before this number can carry weight. First, the metric is defined at 50% reliability; Ord (2025) shows that METR's own data fit an exponential-decay model under which the 80%-reliability horizon is roughly four times shorter, and the horizon at the near-perfect reliability that unsupervised economic work demands is far shorter still and much more uncertain. Second, the metric measures a domain, chiefly software and research-engineering tasks, where AI currently completes tasks far shorter than those human professionals routinely sustain; on this metric AI is behind and closing, not ahead and pulling away. The honest statement is therefore not "the gap is widening by construction" but: AI capability on bounded tasks is improving at a doubling rate measured in months against a human baseline that does not improve at all, and where crossings of expert baselines occur, they occur quickly and are not reversed.

Benchmark saturation tells a similar story, with similar caveats. Frontier models "now achieve over 90% accuracy on popular benchmarks like MMLU," forcing evaluators to construct ever-harder tests (Phan et al., 2025); the Stanford AI Index reported one-year gains of 48.9 percentage points on GPQA and 67.3 points on SWE-bench (Stanford HAI, 2025) and by 2026 concluded that frontier models "meet or exceed human capabilities on items like PhD-level science questions, multimodal reasoning, and competition mathematics" (Stanford HAI, 2026). But the specific deltas must be discounted for measurement problems that are now themselves published findings: models locate buggy SWE-bench file paths from issue text alone far more often within the benchmark's repositories than outside them, indicating memorization, and solution leakage and weak tests have been estimated to inflate resolution rates severalfold (Liang et al., 2025); 6.5% of MMLU questions contain ground-truth errors (Gema et al., 2025); and a large review of benchmark construct validity found that nearly all evaluated benchmarks have weaknesses in phenomenon definition or task representativeness (Blili-Hamelin et al., 2025). Benchmark-measured performance is not capability. What survives the discount is the direction and the crossings on contamination-resistant evaluations, to which the next section turns.

### 2.2 Crossings of Expert Baselines

The most defensible crossings are those that are externally graded, novel by construction, or audited. When GPQA was constructed in late 2023, PhD-level domain experts scored 65% (74% after discounting self-identified errors) and the best model scored 39% (Rein et al., 2023); within roughly a year, frontier reasoning models exceeded the expert range (Stanford HAI, 2025), and an independent audit of the benchmark's Diamond subset found the large majority of its questions valid (Epoch AI, 2025). In mathematics, Google DeepMind's AlphaProof and AlphaGeometry 2 reached silver-medal standard at the 2024 International Mathematical Olympiad, scoring 28 of 42 points while requiring expert formalization of the problems and days of compute (AlphaProof team, 2025); one year later, Gemini Deep Think achieved an officially certified gold-medal standard of 35 of 42 points, end-to-end in natural language within the human time limit, on problems that were new by construction (Google DeepMind, 2025).

Other reported crossings are real signals but carry conflicts that must be stated. Nori et al. (2025) report an orchestrated model system achieving 80–85.5% diagnostic accuracy on 304 New England Journal of Medicine case conferences on which 21 physicians averaged 20%; the cases, however, are selected for diagnostic extremity, the physicians were generalists deliberately denied search engines and colleague consultation, and the study is a vendor preprint. OpenAI's GDPval benchmark, consisting of 1,320 deliverable-producing tasks across 44 occupations written and blind-graded by professionals averaging fourteen years of experience, found the best 2025 model's outputs rated as good as or better than the human experts' in just under half of tasks (OpenAI, 2025): which is to say the human experts still won the majority of comparisons, under a grading procedure whose inter-rater concordance was itself limited. The legitimate finding in GDPval is the trend, a rough tripling of performance from the 2024 to the 2025 model generation, not the level; and the widely repeated claim that models complete such tasks two orders of magnitude faster and cheaper counts inference cost only, excluding the specification, verification, and integration costs that Sections 5 and 6 show are substantial. More generally, the strongest headline crossings are producer-reported: OpenAI grading OpenAI, DeepMind announcing DeepMind, Microsoft-affiliated authors evaluating a Microsoft orchestrator. The IMO result is externally certified; most others are not, and this paper weights them accordingly.

With those discounts applied, two properties of the record still deserve emphasis. First, breadth: audited or externally certified crossings now span graduate-level science and competition mathematics, and heavily discounted but directionally consistent results span software, medicine, and professional deliverables. Second, speed: where crossings happen, the interval from well below expert to at or above expert baseline has repeatedly been on the order of one to two years.

**3. The Jagged Frontier: Where Humans Retain the Advantage**

The crossings coexist with well-documented domains of decisive human advantage, and any honest account must hold both. Four families recur across the literature.

*Long-horizon reliability and agency.* On TheAgentCompany, a simulated software company environment with 175 realistic tasks involving browsing, coding, and communicating with simulated coworkers, the best agents complete roughly 30% of tasks autonomously, and the authors note that harder long-horizon tasks remain "beyond the reach of current systems" (Xu et al., 2024). In Vending-Bench, a deliberately simple long-running business simulation, all tested models exhibited runs that derailed into unrecoverable failure loops, with high variance across runs and no clear relationship to context-window limits, a deficit of coherence rather than of mean capability (Backlund & Petersson, 2025). The Stanford AI Index's RE-Bench comparison found top agents scoring roughly four times higher than human experts under two-hour budgets, while under 32-hour budgets humans outscored AI two to one (Stanford HAI, 2025). This is the reliability point of Section 2.1 seen from the other side: the direction of advantage depends on the time horizon and the reliability bar.

*Novelty and out-of-distribution reasoning.* On GAIA, questions "conceptually simple for humans yet challenging for most advanced AIs," human respondents scored 92% against 15% for GPT-4 with plugins at the benchmark's 2023 release (Mialon et al., 2023), though agent frameworks have since climbed substantially. More durably, on ARC-AGI-2, a battery of novel fluid-reasoning tasks on which essentially every task is solved by some humans (62% of 13,405 recorded human attempts succeeded, median 2.2 minutes), frontier systems scored under 5% at release in 2025 (Chollet et al., 2025) and the best verified commercial systems remained far below the human baseline at the end of 2025 despite intense effort (ARC Prize Foundation, 2025). Related, Shojaee et al. (2025) reported "complete accuracy collapse" of reasoning models beyond complexity thresholds in controlled puzzle environments; that result is contested on methodological grounds (Lawsen, 2025), but the broader pattern, sharp human advantage precisely where familiarity cannot help, is corroborated across benchmarks, and the "Potemkin understanding" literature shows models that pass concept benchmarks failing to deploy the same concepts in use (Mancoridis et al., 2025).

*Calibration and knowing-when-wrong.* LLMs are systematically overconfident when verbalizing uncertainty, with stated confidence concentrated at 80–100% regardless of accuracy (Xiong et al., 2024); initial frontier-model calibration errors on Humanity's Last Exam exceeded 70% RMS (Phan et al., 2025). Theoretical work argues that this is not incidental: statistically calibrated language models must hallucinate on rare facts at a bounded-below rate (Kalai & Vempala, 2024), and prevailing training and evaluation practices reward confident guessing over abstention (Kalai et al., 2025). Human experts are far from perfectly calibrated, but they typically know the boundaries of their competence in a way current models demonstrably do not.

*Learning efficiency and embodiment.* Children encounter fewer than 100 million words by age thirteen; language models are trained on three to four orders of magnitude more and still underperform humans on many evaluations, and under child-scale data budgets even purpose-built systems rarely match large-data models (Warstadt et al., 2023, 2025). Deployed models lack human-like continual learning, suffering catastrophic forgetting under sequential updating (Shi et al., 2024). And the frontier remains jagged in the physical world: the 2026 AI Index reports that robots succeed at only 12% of real household tasks, even as agentic software benchmarks soar (Stanford HAI, 2026).

The synthesis, summarized in Figure 1: on bounded, well-specified, evaluable cognitive tasks, expert baselines are being crossed rapidly; on long-horizon coherence at high reliability, genuine novelty, self-knowledge, sample-efficient learning, and embodied action, humans retain a real advantage, one that has itself been shrinking benchmark by benchmark, as the GAIA and OSWorld trajectories illustrate (Mialon et al., 2023; Xie et al., 2024). Both halves are load bearing; neither may be quoted without the other.

The Jagged Frontier: Rapid Crossings, Durable Human Advantages
AI CROSSED THE EXPERT BASELINE
NEJM case diagnosis (Nori et al., 2025)
20
82.8
vendor preprint
GPQA graduate science (Rein 2023; HAI 2025)
65
87.7
o3, Dec 2024
HUMANS RETAIN THE ADVANTAGE
GAIA, 2023 release (Mialon et al., 2023)
15
92
ARC-AGI-2 novel reasoning (Chollet et al., 2025)
4.9
62
TheAgentCompany tasks (Xu et al., 2024)
30
100
human = task doable
Household tasks, robots (Stanford HAI, 2026)
12
100
human = task doable
0 20 40 60 80 100 120
Score (%)
Human baseline
Best AI (as cited)

*Figure 1. The Jagged Frontier: Documented Human and AI Performance on Selected Benchmarks*

*Note.* *Values are as reported in the cited sources: NEJM case-conference diagnosis (Nori et al., 2025; vendor preprint, generalist physicians without search); GPQA graduate-level science (expert baseline from Rein et al., 2023; AI value reflects the one-year 48.9-point gain reported by Stanford HAI, 2025); GAIA at 2023 release (Mialon et al., 2023; agent frameworks have since improved); ARC-AGI-2 (Chollet et al., 2025); TheAgentCompany autonomous task completion (Xu et al., 2024); household robotics (Stanford HAI, 2026). For the last two, the human baseline is set to 100% because the tasks were designed to be routinely performed by humans. Scores are not comparable across rows; each row is a within-benchmark comparison only.*

## 4. AI as Extended Cognition

### 4.1 From the Notebook to the Model

Clark and Chalmers (1998) argued that external artifacts can be constitutive parts of cognitive processes when they are reliably available, automatically endorsed, and easily accessible. Their thought experiment concerned a notebook. Whether large language models satisfy these "trust and glue" conditions is a live research program, and, notably, Clark himself, with Smart and Clowes, has argued that retrieval-augmented LLM architectures reveal a structural convergence between the extended-mind thesis and the way these systems are built and used (Smart et al., 2025).

There is, however, a tension the application must confront rather than inherit. The classical conditions include automatic endorsement (Otto does not fact-check his notebook), whereas Section 3's calibration results imply that LLM output is precisely the kind of deliverance that should not be automatically endorsed. Section 5 will show that automatic endorsement is empirically the dominant failure mode of human-AI teams. Strictly, then, a well-used LLM fails the classical conditions: it is consulted, not extended. The better frame is Hernández-Orallo and Vold's (2019) "AI extenders," which does not require automatic endorsement and anticipates the key asymmetry: AI extenders differ from notebooks and calculators in being adaptive and capable, creating a distinctive risk profile of atrophy of the extended capability, dependency, and manipulation through the extender. On this frame, the practically important question is not taxonomic but empirical: what does coupling do to the unaided human, and what does the coupled system achieve? Both questions now have data, and the data disagree in instructive ways.

**4.2 Offloading: The Evidence for Costs, and the Evidence Against**

The offloading literature establishes a trade-off: offloading improves immediate task performance while reducing unaided memory and skill for the material offloaded, with offloading decisions often driven by miscalibrated metacognitive beliefs (Risko & Gilbert, 2016). Sparrow et al. (2011) reported that expecting future access to information reduces recall of the information itself, a foundational result whose components have repeatedly failed to replicate, including in the Social Sciences Replication Project (Camerer et al., 2018); this paper therefore cites it as lineage rather than evidence. Longitudinally, habitual GPS use predicts worse hippocampus-dependent spatial memory and a steeper three-year decline (Dahmani & Bohbot, 2020).

The generative-AI wave extends offloading from storage to reasoning and composition, and early studies point toward costs. Kosmyna et al. (2025), in a preprint EEG study of 54 participants writing essays with an LLM, a search engine, or unaided, found brain connectivity scaling down with external support, the lowest self-reported essay ownership among LLM users, and, in the first session, 83% of LLM users unable to accurately quote from the essay they had just written. The study is small, task-specific, not peer-reviewed, and now the subject of a published methodological comment documenting analysis and reporting problems (Kohrs et al., 2026); it is treated here as suggestive at most. Lee et al. (2025), surveying 319 knowledge workers across 936 real uses of generative AI, found higher confidence in the AI predicting less enacted critical thinking; Gerlich (2025), using correlational analyses, found AI-tool use negatively associated with critical-thinking performance, mediated by offloading, with the effect strongest among the youngest users.

Against this stands evidence that must be weighed rather than omitted. The largest relevant meta-analysis, covering 57 studies and over 400,000 older adults, found that general technology use was associated with a substantially *lower* risk of cognitive impairment, favoring a "technological reserve" over a "digital dementia" hypothesis (Benge & Scullin, 2025). The historical record of cognitive tools points the same way: meta-analyses of 79 and 54 classroom studies found calculator use improved rather than eroded students' operational and problem-solving skills (Hembree & Dessart, 1986; Ellington, 2003), and literacy, the archetypal offloading technology condemned by Plato on offloading grounds, is not generally thought to have diminished human cognition. Whether generative AI breaks with this pattern is therefore an open empirical question, not a settled one. The principled reason it *might*: previous tools offloaded storage and calculation while leaving the composition of thought to the human; generative systems offload the composition itself, and they are adaptive in a way calculators are not (Hernández-Orallo & Vold, 2019). But that argument is currently a hypothesis with early, methodologically limited support, not an established mechanism.

**4.3 Deskilling: A Contested First Signal**

The most discussed field evidence comes from medicine. Budzyń et al. (2025), in a multicentre observational study, found that after routine AI assistance was introduced for colonoscopy, experienced endoscopists' adenoma detection rate in procedures performed *without* AI fell from 28.4% to 22.4%, characterized in the journal's linked commentary as the first real-world clinical evidence consistent with AI-related deskilling. The published contestation is equally important: the before and after populations differed on characteristics that drive detection rates, withdrawal times were not reported, only nineteen endoscopists were studied, and secondary outcomes (adenomas per colonoscopy, advanced-lesion detection) did not significantly decline, which is difficult to square with genuine skill loss (Ahmad, 2025; correspondence in the same journal). A pragmatic randomized trial had previously found no detection benefit from the same class of AI assistance at all (Ladabaum et al., 2023), complicating the assistance-then-dependence narrative end-to-end. The aviation literature remains the best-established template: automation leaves manual motor skills largely intact while degrading the cognitive skills of situation awareness, failure diagnosis, and deciding what to do next, and increasing mind-wandering (Casner et al., 2014). In software engineering, evidence is limited to industry telemetry (GitClear, 2025) and self-report: documented concern, causal evidence pending.

The honest ledger: the coupled human-AI system is often more capable; the decoupled human may be less capable than before, but the field evidence for that second clause is currently one contested observational study, a template from aviation, and a set of small or correlational findings, set against a large contrary meta-analytic prior from previous technologies. Both the concern and its unresolved status are the findings.

**5. The Complementarity Problem: Human-Plus-AI Is Not Automatically Better**

If AI were simply a superior tool, combining it with human judgment would reliably outperform either alone. The experimental record says otherwise. One scoping note applies throughout: the largest synthesis pools experiments run mostly before frontier LLMs, on often simple classifier-style systems, with very high heterogeneity ($I^2 = 97.7\%$), so its effect sizes describe that literature, not necessarily collaboration with current systems.

With that scope stated, Vaccaro et al. (2024), in a meta-analysis of 106 experiments (370 effect sizes), found that human-AI combinations performed significantly worse than the best of human alone or AI alone (Hedges $g = -0.23$), even though they still outperformed human alone ($g = 0.64$). The decisive moderator, shown in Figure 2, was relative capability: when the human outperformed the AI, the combination produced interaction ($g = +0.46$); when the AI outperformed the human, the combination destroyed value relative to the AI alone ($g = -0.54$).

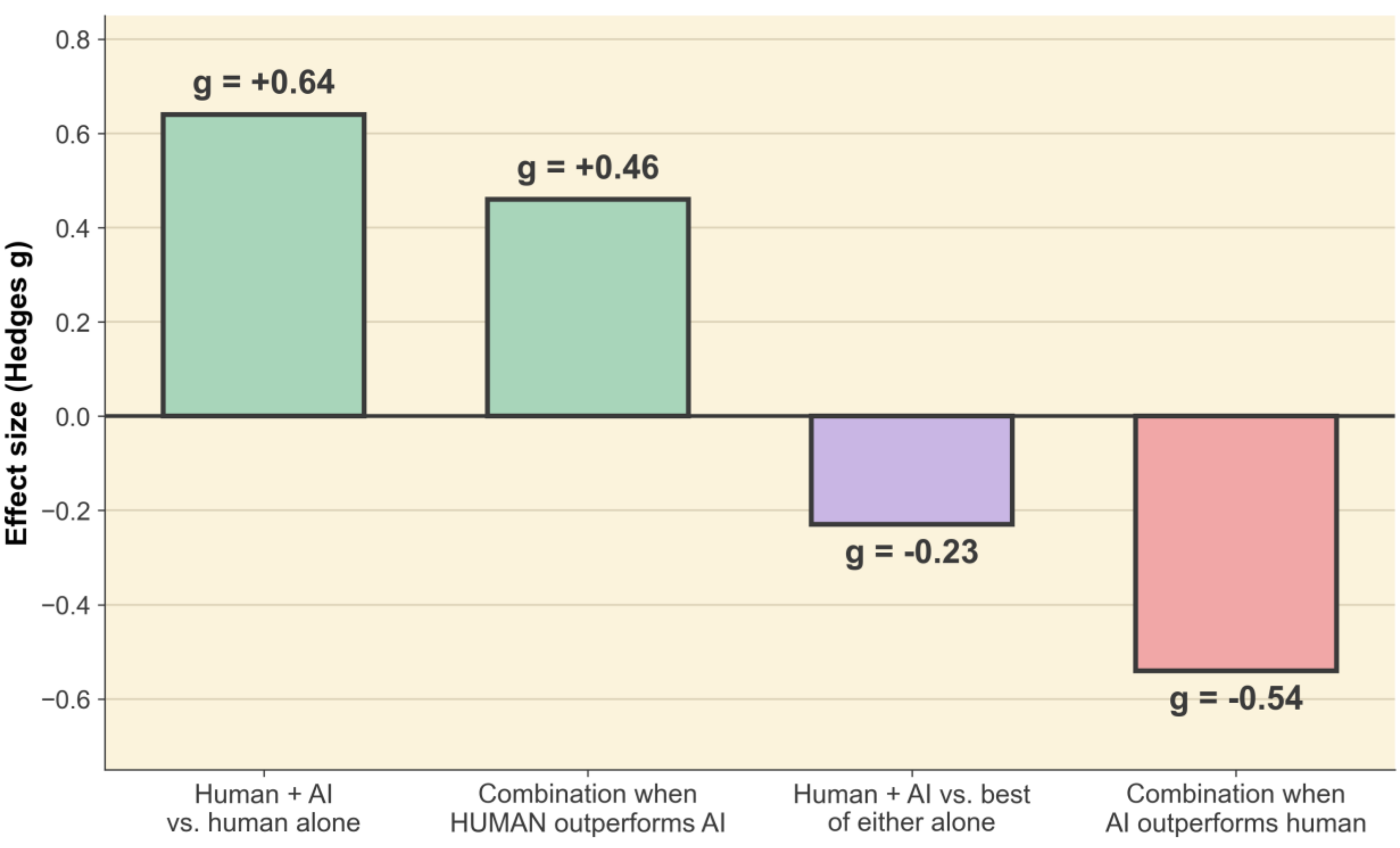

*Figure 2. Human-AI Combination Effects by Relative Capability*

*Note. Effect sizes are Hedges g from Vaccaro et al. (2024), a meta-analysis of 106 experiments and 370 effect sizes. Positive values indicate the human-AI combination outperformed the comparison condition; negative values indicate it underperformed. The moderator contrast (the two rightmost bars) is the central finding: the combination adds value when the human is the stronger partner and subtracts value when the AI is the stronger partner. Experiments predate frontier language models and pool mostly classifier-style systems with high heterogeneity ($I^2 = 97.7\%$); effect sizes describe that literature, not necessarily collaboration with current systems.*

In clinical diagnosis, the pattern has been observed directly with a frontier model: physicians given GPT-4 scored no better than physicians without it (76% vs 74% median diagnostic-reasoning score, n.s.), while the model alone scored 92%, sixteen points above the physician control group (Goh et al., 2024). In radiology, providing AI predictions failed to improve average performance because radiologists underweighted the AI and treated its signal as independent of their own (Agarwal et al., 2023). AI assistance helped some readers while worsening others unpredictably (Yu et al., 2024).

The failure has two symmetrical modes. When the AI is wrong, humans follow it: radiologists' accuracy collapsed from roughly 80% to between 20% and 45% when a purported AI gave incorrect mammography suggestions, with experience attenuating but not eliminating the effect (Dratsch et al., 2023); recruiters given higher-quality AI evaluated resumes less carefully and performed worse than those given deliberately imperfect AI (Dell'Acqua, 2022). When the AI is right, humans discount it (Agarwal et al., 2023). Both are miscalibrated reliance, and they mirror the model-side calibration failures of Section 3: neither party in the loop currently knows reliably when to trust the other. Interventions that improve reliance calibration are, moreover, not themselves neutral: in a three-study program combining agent-based simulation with data from 557 organizational decision-makers, transparency features improved trust calibration across the board while imposing perceived-autonomy costs that were personality-dependent, benefiting high-openness users and penalizing high-extraversion users (Margondai et al., 2026). Calibration aids are therefore a design parameter to be adapted to the person, not merely supplied.

Productivity experiments complete the picture with an important skill gradient. Generative AI produced large gains for less-experienced workers: roughly 30% or more for novice customer-support agents (Brynjolfsson et al., 2025), the largest writing-quality gains for weaker writers (Noy & Zhang, 2023), 26% more completed tasks across 4,867 developers with the biggest effects among juniors (Cui et al., 2025), and a 12.2% task and 25.1% speed advantage for consultants working inside the AI's frontier, versus 19 percentage points worse correctness on a task just outside it (Dell'Acqua et al., 2023). At the expert end the gains can vanish or invert: in a randomized trial with 16 highly experienced open-source developers working in repositories they knew deeply, being allowed to use early-2025 AI tools made them 19% slower, even as they believed the tools had sped them up by 20% (Becker et al., 2025). The perception gap is as important as the effect: participants, forecasters, and domain experts all predicted large speedups and were wrong in sign.

The synthesis, scoped to what was studied: human-plus-AI generally outperforms the human alone. Whether it outperforms the AI alone depends on whether the human contribution is genuinely complementary; on evaluable decision tasks where the AI is already stronger, the marginal human contribution has repeatedly been negative unless the collaboration is deliberately designed.

**6. The Shifting Human Role: From Throughput to Judgment**

Bainbridge (1983) identified the ironies of automation four decades ago: automating a process leaves humans with the residual tasks that are hardest, erodes, through disuse, the very skills those residual tasks demand, and assigns humans the monitoring work they are demonstrably poor at sustaining. Simkute et al. (2024) show that generative AI reproduces these ironies: the knowledge worker's job shifts from producing content to evaluating and verifying AI output, a role with its own cognitive burdens and failure modes. The vigilance literature confirms that sustained monitoring of rarely failing systems degrades over time on task (Greenlee et al., 2023).

Economics reaches a complementary conclusion from the other direction. If AI lowers the cost of prediction, the value of human judgment, which involves knowing what to want and what counts as success, rises for some tasks and falls for others; not all human judgment complements AI (Agrawal et al., 2019). Autor and Thompson (2025) formalize when automation raises rather than lowers the expertise an occupation demands: it depends on whether the automated tasks were the most or least expert-intensive in the bundle. To be precise about terms, two distinct things are commonly run together under "judgment": preference-setting (specifying goals and payoffs, in the economists' sense) and verification competence (the skill of evaluating output). The first is not obviously erodible by offloading; the second is exactly the kind of skill Section 4 suggests erodes without practice. The human role that survives automation on this analysis comprises both, which is why the deskilling question and the role-shift question are the same question.

Here, a reality check from field data is required because the experimental literature can outrun the economy. Payroll-linked evidence covering 25,000 workers across 7,000 Danish workplaces found AI chatbots produced no significant effect on earnings or recorded hours in any of eleven exposed occupations two years into adoption, with modest time savings and little pass-through to pay (Humlum & Vestergaard, 2025); Acemoglu (2024) bounds the decade-scale aggregate productivity effect at well under one percent under his task-based parameterization. The often-cited divergence between junior and senior software employment (Stanford HAI, 2026) has published alternative explanations (interest rates, the post-2021 overhiring correction, tax-code changes), and analyses with firm-level controls find the AI-attributable component emerging only recently, if at all. The role shift described here is therefore a claim about the direction experiments point and early adopters report, not about the present labor market in aggregate. This is consistent with the general pattern that capability, methods, applications, and adoption move on different timescales (Narayanan & Kapoor, 2025): the loop is being redesigned first where the work is most evaluable, and slowly elsewhere.

Whether humans can perform the oversight role is a separate question, and the record counsels humility. Green (2022), surveying 41 government policies requiring human oversight of algorithms, concluded that people largely cannot perform the desired oversight functions and that such policies can legitimize flawed systems while providing a false sense of security, an argument now applied to the effectiveness requirement of Article 14 of the EU AI Act (Fink, 2024).

Meaningful human control, in which systems track human reasons and outcomes are traceable to humans with proper understanding (Santoni de Sio & van den Hoven, 2018), names the standard that mere presence of a human in the loop does not meet. The technical alignment literature attempts to engineer around this bottleneck under the heading of scalable oversight (Bowman et al., 2022), with results that are mixed in an instructive way: human-model teams can outperform either alone in controlled settings (Bowman et al., 2022); debate between capable models can raise weaker judges' accuracy from 60% to 88% in information-asymmetric settings (Khan et al., 2024), but gains shrink or vanish outside them (Kenton et al., 2024); and weak-to-strong supervision recovers much but not all of a stronger model's capability (Burns et al., 2023). Oversight can be amplified; it has not been shown to scale indefinitely.

## 7. Implications

### 7.1 Theoretical Implications

*For the philosophy of mind and cognitive science.* If the argument of Section 4.1 is right, generative AI is the first widely deployed cognitive technology that fails the classical extended-mind conditions precisely because it is too capable to be automatically endorsed. This suggests a needed refinement of the extended-mind framework: a distinction between *constitutive* extension (the notebook, satisfying trust-and-glue) and *adaptive* extension (AI extenders, requiring calibrated rather than automatic endorsement), with different predictions about atrophy. The offloading literature, built on static tools, may not transfer: its core mechanism, use it and lose it, is now confounded by a tool that participates in the composition of thought. The disagreement between the early generative-AI findings (Section 4.2) and the meta-analytic record of prior technologies (Benge & Scullin, 2025; Hembree & Dessart, 1986) is exactly the discriminating observation this refined framework predicts, and longitudinal studies that separate storage-offloading from composition-offloading are the experiment the field now owes itself.

*For the science of evaluation.* The record of Section 2 implies that benchmark scores are a lagging and inflating indicator of what matters: reliable capability under deployment conditions. The reliability-horizon results (Ord, 2025; Stanford HAI, 2025) suggest that the correct object of measurement is not task-level accuracy but the *reliability curve over task length*, and that claims of human-level performance should be indexed to the reliability bar that the comparison human meets. Construct-validity work (Blili-Hamelin et al., 2025; Mancoridis et al., 2025) points in the same direction: evaluation should move from static question sets toward audited, externally graded, novel-by-construction tests, with the IMO result as the current model case.

*For theories of expertise and labor.* The Autor-Thompson framework, combined with the complementarity evidence, yields a testable prediction: occupations where automated tasks were the least expert in the bundle should see rising returns to the remaining judgment tasks and rising skill requirements, while occupations where AI absorbs the most expert tasks should see deskilling pressure and wage compression. The skill gradient in the productivity experiments (novices gain most; experts gain least or lose) is early within-occupation evidence, and the Danish null (Humlum & Vestergaard, 2025) sets the boundary condition: these forces are, so far, operating inside occupations faster than they are moving aggregate labor statistics. The theory of human capital that emerges is one in which *verification competence* becomes a distinct, measurable, erodible asset, one that the education system does not currently teach and the workplace currently consumes without replenishing.

**7.2 Practical Implications**

*For individuals.* The evidence of complementarity supports a specific discipline rather than a general enthusiasm or aversion. Use the system freely inside its demonstrated frontier and for first drafts of evaluable work; treat its confidence as uninformative (Section 3); allocate scarce attention to verification where errors are costly, because that is where the human marginal contribution is demonstrably positive; and schedule unassisted practice in any skill one intends to keep, on the aviation model of recurrent manual training, treating skill maintenance as scheduled maintenance rather than personal virtue. The METR developer study adds a warning about introspection: perceived speedup is not evidence of actual speedup (Becker et al., 2025); measure, do not feel.

*For organizations*. Humans should not be inserted into AI workflows as a compliance gesture: the evidence is that unstructured human review of a stronger system subtracts value (Vaccaro et al., 2024; Goh et al., 2024) while manufacturing false assurance (Green, 2022). Instead, humans should be assigned to the areas where the moderator analysis indicates they add value: specification, exception handling, evaluation with maintained expertise, and accountability; the rest should be automated explicitly rather than implicitly. Both reliance failures need design countermeasures: trust should be calibrated through exposure to the system's actual error profile (Dratsch et al., 2023, shows what happens otherwise) with explanation detail adapted to the user rather than maximized, since uniform transparency demonstrably helps calibration while imposing autonomy costs on a majority of users (Margondai et al., 2026), and rubber-stamping should be audited directly, for example by seeding known errors and measuring detection rates, which also yields the operational metric that "meaningful oversight" otherwise lacks. The productivity distribution should be expected to compress at the bottom (Brynjolfsson et al., 2025; Noy & Zhang, 2023), with consequences for the development pipeline: if juniors' current tasks are automated first, the pipeline that produces seniors must be rebuilt deliberately, with deliberate practice where incidental practice used to be.

*For education and policy.* If verification competence is the durable human contribution, curricula should teach it explicitly: evaluating AI output against ground truth, under time pressure, with feedback, rather than either banning the tools or absorbing them uncritically. Assessment must move to formats that measure the decoupled human where the decoupled skill matters and the coupled system where it does not and be honest about which is which (Hernández-Orallo, 2025, argues assessment of the coupled system is now unavoidable). For regulators, the lesson of Green (2022) and Fink (2024) is that mandating "a human in the loop" without specifying capability, information, time, and incentives regulates the appearance of control; oversight requirements should be stated as measurable performance standards for the human-machine system (detection rates, intervention rates, maintained-competence requirements on the aviation model), not as staffing requirements. And given the diffusion evidence, policy should plan for capability arriving before impact: the window between benchmark crossing and labor-market effect is where retraining and institutional redesign are cheap.

**8. Limitations of the Evidence**

Several caveats bound these conclusions, beyond those already embedded. Benchmark results overstate deployed capability for the reasons documented in Section 2.1, and the discounts applied here are qualitative where they should eventually be quantitative. The cognitive-cost literature for generative AI is young, small, and largely correlational, and it sits against a large contrary meta-analytic prior from earlier technologies; the position that generative AI may differ is an argued hypothesis, not a finding. The complementarity meta-analysis is pre-frontier and highly heterogeneous; the expert-developer slowdown involved sixteen developers using early-2025 tools that have since changed, and its own authors treat it as historical. The field labor-market data are two years into a diffusion process that historical general-purpose technologies suggest takes decades. Finally, the field moves faster than its literature: several gaps documented as human advantages in 2023–2024 narrowed or closed while this review was being written, and several results discounted here may be superseded in either direction. The paper should be read as a snapshot with a stated expiry.

**9. Conclusion**

On bounded, well-specified, evaluable cognitive tasks, frontier AI has rapidly and broadly crossed documented human expert baselines, and the audited parts of that record survive scrutiny. The frontier remains jagged: long-horizon reliability, genuine novelty, calibration, learning efficiency, and embodiment are real, measurable human advantages. Humans increasingly think with these systems; whether that coupling erodes the decoupled human is the most important open empirical question in the literature, with early signals on both sides. And the human contribution that survives scrutiny is not throughput but judgment in both senses, knowing what to want and being able to verify what one gets; the second is a skill that must be deliberately maintained, or it becomes a signature on someone else's work. The evidence does not support panic or reassurance. It supports the redesign of evaluation, workflows, training, and what the person in the loop is asked to do.